# Testing the Dynamic Time Warping algorithm in the Seafloor Spreading Rate Problem

Ivanov S.A.[1], Merkuriev S.A.[1,2]

*1.-Marine Geomagnetic Investigation Laboratory, SPbF IZMIRAN, St.Petersburg, Russia, e-mail: Sergei.A.Ivanov@mail.ru*

*2.-Institute of Earth Sciences, St.Petersburg State University, St.Petersburg, Russia;*

## 1. Introduction and Background.

Information about spreading rate variation over time is one of the most important characteristics of seafloor spreading for understanding of driving forces of plate motions. Different parts of the global Mid-Ocean Ridge system create oceanic crust at rates that differ by more than a factor of ten. Even in one region the spreading rate can be changed by 50%. An efficient way to understand the geodynamical mechanism of oceanic lithosphere accretion is the detailed observation of variations in space and time of the accretion rate [Le Pichon et al., 1973]. The knowledge of the spreading rate gives information about plate movement and also about the production of new oceanic crust [Purdy and all 1992].

There are two ways to determine relative spreading rates. The first one consists of determination of the distance between two isochrones of different ages. The second method is more precise but more time-consuming because it requires calculating the poles of rotation for the pair of isochrones of the same age, what gives the average spreading rate on this time interval. To find the interval spreading rate we need to repeat this procedure for the whole sequence of isochrones and then obtain the stage poles of rotation [Cox and Hart, 1986]. Evidently, that the both methods allow us to determine only the average spreading rates for the time intervals between chrones. The temporal resolution of the spreading rate variation usually is determined by the spatial density of marine magnetic anomaly identifications. Usually the value of the time interval is 5-10 Ma, but last time the resolution of temporal spreading rate variation increases up to 1 Ma [Merkouriev and DeMets, 2008] or smaller [Weiland, 1995].

Interval spreading rate variation with time obtained by using the traditional method of identification gives a piecewise constant function. However it is well known that during evolution of oceanic basin the temporal behavior of the spreading rate can be more complicated, for examples, the spreading can be stopped or can increase linearly. Thus, the study of these types of non-stationarity of the spreading process using traditional technique with a linear correlation is useless and require of application special algorithm.

We offer to apply the powerful Dynamic Time Warping (DTW) algorithm to find the spreading rate variation by comparing profiles of marine magnetic anomalies with the synthetic field constructed by the magnetic polarity reference scale where polarity reversals are well dated via other independent methods. This approach was used in [Lallier et all, 2013] for absolute dating of sediments where a measured polarity column in a sediment section is compared with the geomagnetic polarity time scale. In this work two binary sequences are compared (two telegraph signals). In contrast to this we are going to use the DTW algorithm comparing the observed profile (or a stack) of a marine magnetic anomaly with the synthetic profile calculated using the geomagnetic polarity timescale.

First the method similar to the DTW, where aligning synthetic and observed profile on the basis of syntactic recognition and calculation of Levenstein distance, was applied to obtain the age grid of the crust at the Mid-Atlantic Ridge between 28° and 29°N [Zhizhin et all, 1997 ]. The authors wrote that 'it is possible to show temporal variations in spreading rate, lateral variations between profiles".

Before to apply the DTW approach to real data we show the testing results of the algorithm by the following way. With the known spreading rate and the given interval of the polarity scale we generate the synthetic magnetic fields which will be used as a pattern. Here we consider only classical model Vine-Mathews. This field we compare with the field obtained for temporal variations of seafloor spreading rate. These variations are equivalent to transformation of the magnetic time scale.

It is important to note that the transformation of the scale (e.g., dilatation) does not lead to the same transformation of the field. This is the reason why we are not able to recover the transformation exactly even for the synthetic fields. Nevertheless we hope that the algorithm can be applied to study of plate movements.

## 2. The method and the results

We use a DTW algorithm which minimizes the effects of gaps, different speed, and other distortions of two `time-series' by transformation in order to detect similar shapes. Let we have two time series X = (x(1), x(2), ...x(N) ), and Y = (y(1), y(2), ...y(M)), represented by the sequences of vertices of two curves. The data sequences should be sampled at equidistant points in time. Note that here `time' is a parameter of the curves, in fact we have the length. To use the algorithm we have specify the `cost matrix', i.e. a set of all `distances' between any pairs ( x(n), y(m) ). The output is the path – the sequence of the pairs (x($n_r$), y ($m_r$)) such that the cost of this path is minimal. The path has to be

(i) monotonic by x and by y,

(ii) satisfies the boundary condition: the start is (x(1), y(1)) and the end is (x(N), y(M)).

We choose the simplest step size, namely from the point ( x(m), y (n) ) we can go only to three pairs: ( x(m+1), y (n) ), ( x(m), y (n+1) ), and ( x(m+1), y (n+1) ). It is interesting that the best result corresponds to the "distance" $|x-y|^{1/2}$.

For the testing the algorithm we use geomagnetic time scale [Gradstein, Ogg, Schmitz, 2012], exactly the interval with the last reversal 14.609 Ma. First we find the pattern, namely the magnetic field corresponding to this interval and the Vine-Mathews block model for the half spreading rate 15 km/Ma. The sampling rate is 250 m. Then we calculate the profiles corresponding to the three scale transformation:

(i) The gap – the spread rate is the same, but the reversals from 90.495 Ma to 112.335 Ma are removed.

(ii) The jump of the spreading rate: at the middle of the time interval the spreading rate increases in 1.8 times.

(iii) Acceleration of the spreading rate: at the first half interval the spreading rate is 19 km/Ma. Then at the middle of the time interval the spreading rate linearly decreases and reaches 9 km/Ma.

The figures 1,2, and 3 show the results of the our algorithm for these three tests. All the figures have the same structure: on the left there is the plot of the synthetic field corresponding to unperturbed scale; the plot of the field corresponding to perturbed scale is below; in the center of the figures there are the path corresponding to the scale transformation (red line) and the obtained optimal path (white wide line); shaded area is the cost matrix and its scale is on the right (regions of low cost are indicated by dark colors and regions of high cost are indicated by light colors).

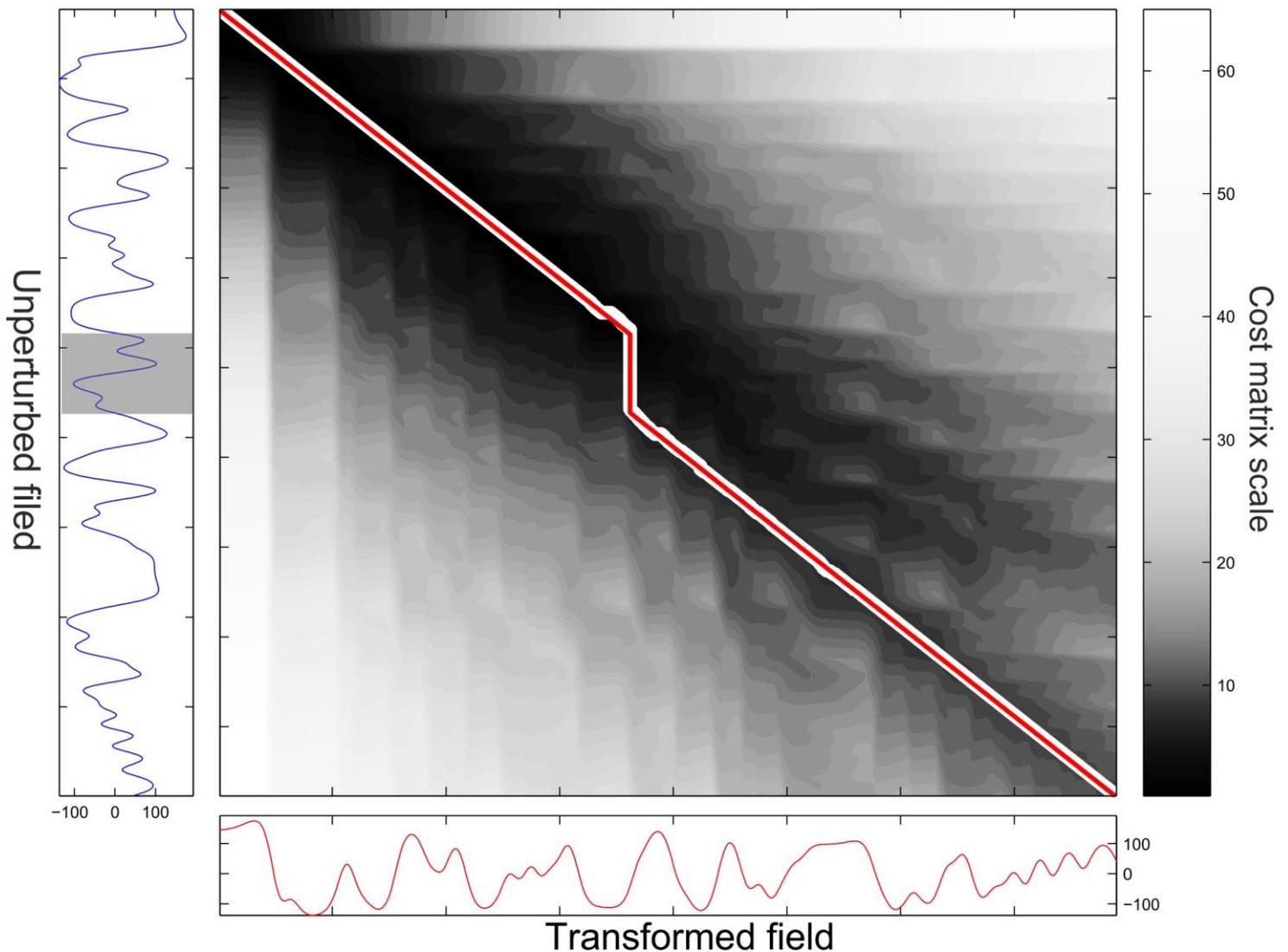

Fig. 1. The case (i). At the left plot the shaded area shows the magnetic anomalies missed in the transformed field.

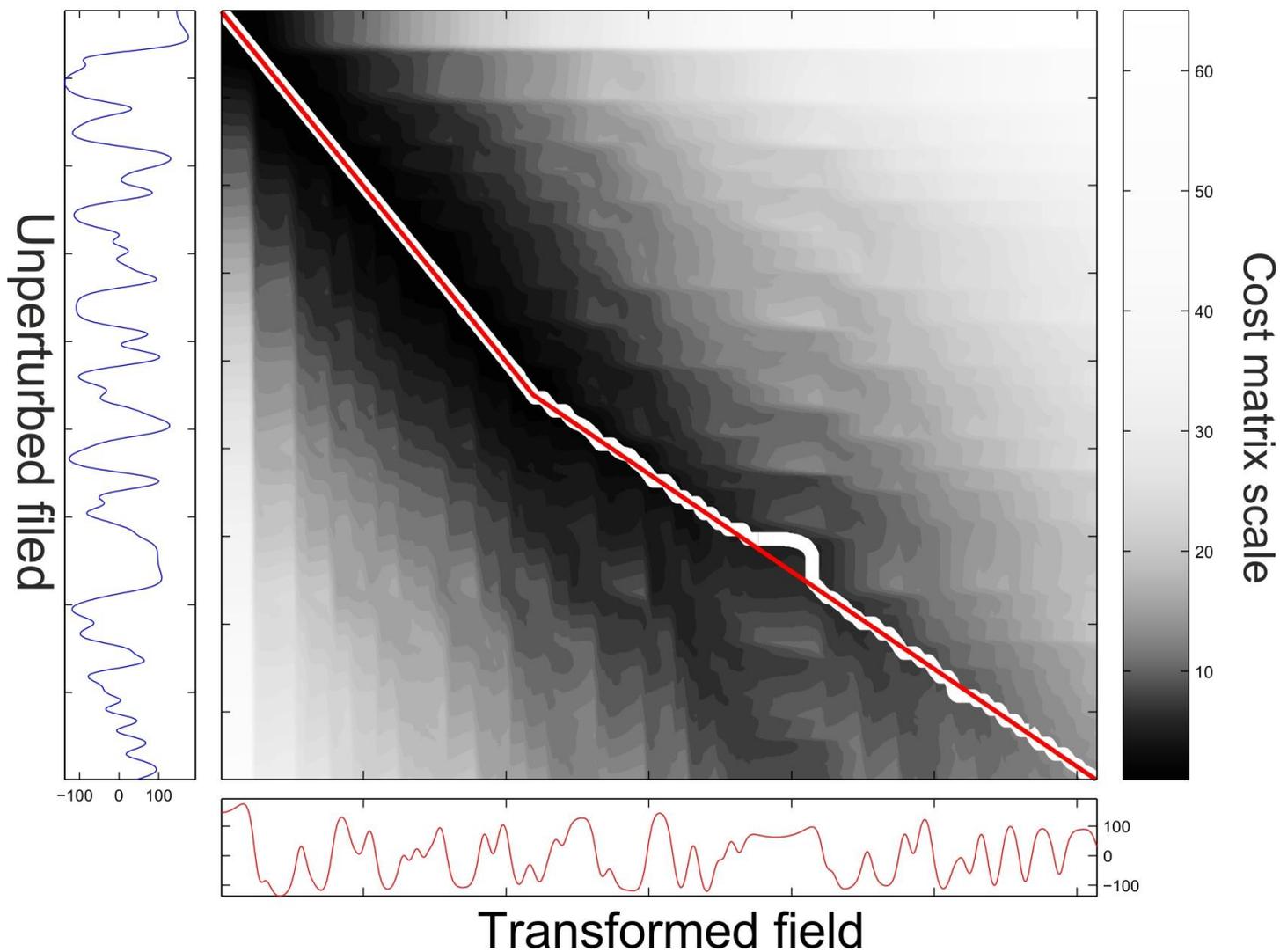

Fig. 2. The case (ii). Note that deviation of the optimal path from the exact path corresponds to the center of the large magnetic anomalies with the low amplitude variation.

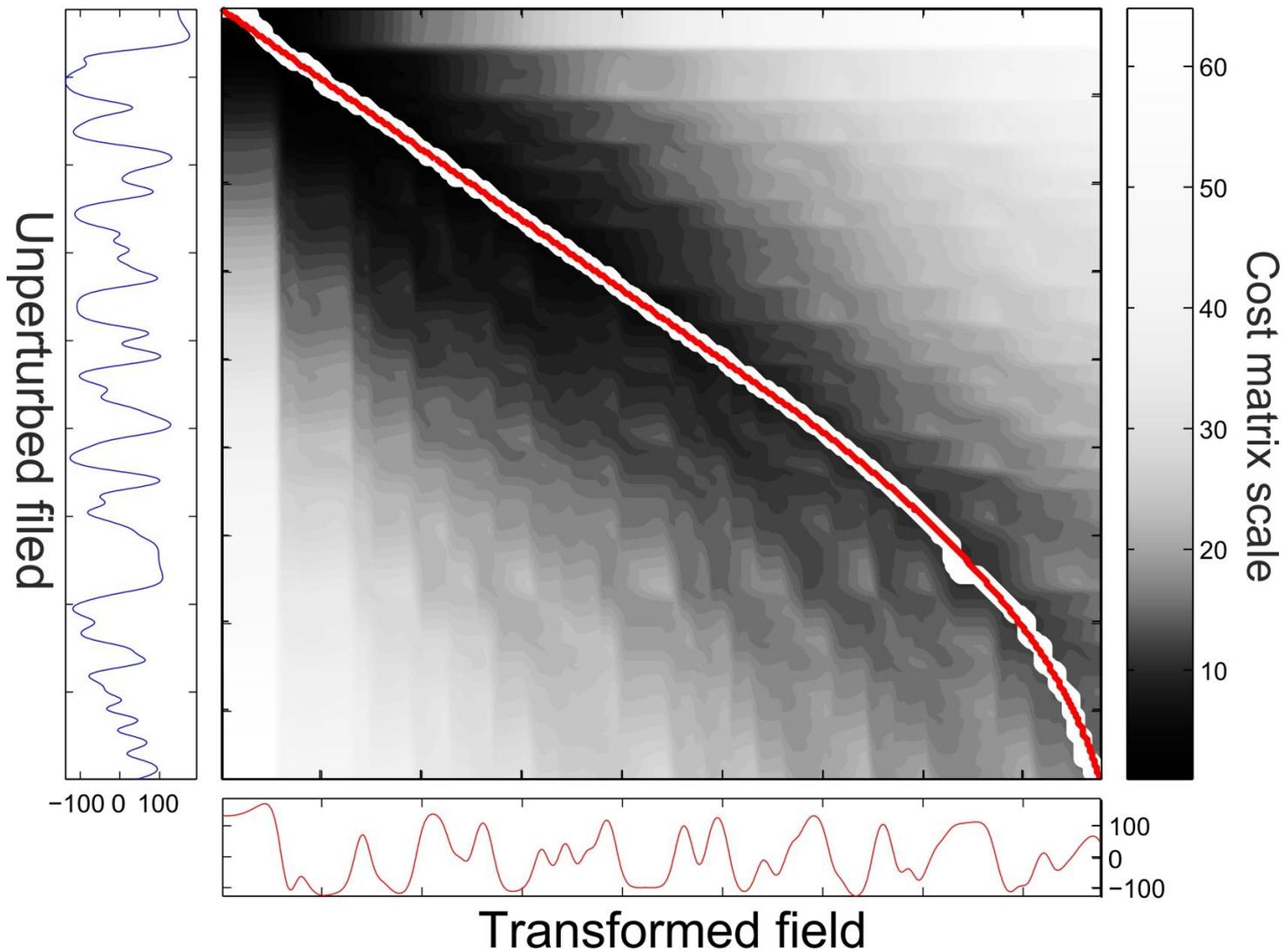

Fig. 3. The case (iii).

### 3. Subsequence DTW

In many applications, the sequences to be compared exhibit a significant difference in length. Instead of aligning these sequences globally, one often has the objective to find a subsequence within the longer sequence that optimally fits the shorter sequence. For example, assuming that the longer sequence represents a model field constructed by a whole scale of reversals and the shorter sequence is a part of the observed anomaly. A typical goal is to identify the fragment within the part of the model field. The problem of finding optimal subsequences can be solved by a variant of dynamic time warping also.

### 4. Conclusion

We show that the DTW method can be applied to find the variations of the spreading rate.

### 5. Acknowledgement

This work was supported by the Russia Foundation of Basic Research grants 15-05-06292.